\documentclass[twoside,slac_one]{revtex4}
\usepackage{graphicx}
\usepackage{fancyhdr}
\usepackage{amsmath} 
\usepackage{bm}
\usepackage{amsxtra}
\usepackage{amssymb}
\usepackage{amsthm}
\usepackage{latexsym}
\usepackage{lscape}

\begin{document}

\title{Project 8: Using Radio Frequencies to Measure the Neutrino Mass}

%

\author{N.\,S.~Oblath}
\affiliation{Laboratory for Nuclear Science, Massachusetts Institute of Technology, Cambridge, MA, USA}

\begin{abstract}
It is well known that the neutrino masses affect the shape of the energy spectrum of tritium beta-decay electrons. However, experiments have yet to measure that distortion. The Project 8 experiment proposes to measure the spectral distortion in a novel way: using radio-frequency techniques to detect and measure the energies of the beta-decay electrons. We plan on measuring the radiation created from the cyclotron motion of the electrons in a strong magnetic field. I will report on the status of a prototype that is designed to demonstrate single-electron detection at energies near the tritium endpoint, 18.6 keV. I will also discuss the possibilities for scaling up to a neutrino-mass experiment.
\end{abstract}

\maketitle

\thispagestyle{fancy}


\section{Introduction}
While neutrino oscillation experiments have successfully shown that neutrinos change flavor, and therefore have non-zero mass, the absolute mass scale remains unknown.  Oscillation experiments have accurately measured two independent mass differences between the three known mass states.

So far the data only provide upper limits on the neutrino mass, but experiments continue to probe lower mass ranges.  If zero-neutrino double-beta-decay is measured, it can be interpreted as a measurement of neutrino mass, provided that the decay occurs via straight-forward weak processes.  Neutrino mass limits can be implied from cosmological data, as well, due to the effects of neutrino mass on structure formation in the universe.  The third method is using single beta decays.  The oscillation data indicate that the neutrino mass measured by beta decay experiments, $m_{\beta\nu}$, must satisfy either $m_{\beta\nu} > 0.005\ \mathrm{eV}$ or $m_{\beta\nu} > 0.05\ \mathrm{eV}$, depending on the ordering of the mass states.

The most successful direct measurements of the neutrino mass to-date have been performed using tritium beta decays.  Neutrino mass has an effect on the kinematics of decay process~\cite{ref:tritiumbetadecay}.  While the neutrinos themselves are difficult to measure, the energy of the outgoing electrons can be precisely determined.  The neutrino mass can then be inferred from the shape of the electron energy spectrum.

The limits on the neutrino mass as measured with tritium beta decays have steadily improved over several generations of experiments.  The next-generation tritium experiment, KATRIN, is currently under construction.  It anticipates having a sensitivity of 200~meV~\cite{ref:katrin,ref:katrin2}.

The lower limits for the neutrino mass from oscillation experiments provide a strong motivation for probing to lower neutrino masses.  However, with KATRIN, the technologies used in spectrometer-type tritium experiments have been pushed to their current practical limits.  A new technique is needed to push the mass limits lower.  The technique proposed is the use of cyclotron radiation emitted by the beta decay electrons to measure their energy spectrum.  To determine an electron's energy from its cyclotron radiation, we plan on measuring the frequency of that radiation; fortunately, the frequency of radiation can be measured very precisely, making the technique an attractive method for a potential future neutrino mass measurement.

\section{Neutrino Mass via Tritium Beta Decay}

The best prospects for directly measuring the mass of the neutrino, and the current limits for such measurements, come from the use of beta decays.  The neutrino mass is probed by carefully measuring the energy of the outgoing electron from the decay.  The energy spectrum for beta-decay electrons is given in Eq.~\ref{eq:betadecayspectrum}:

\begin{equation}
\frac{dN}{dK_e} \propto F(Z,K_{e}) \cdot p_e \cdot (K_{e}+m_{e}) \cdot (E_{0}-K_{e}) \cdot \sum_{i=1}^{3}|U_{ei}|^2 \sqrt{(E_0-K_e)^2-m_i^2} \cdot \Theta(E_0-K_e-m_i).
\label{eq:betadecayspectrum}
\end{equation}

The Fermi function, $F(Z,K_{e})$, takes into account the Coulomb interactions of the electron with the recoiling nucleus; $Z$ is the proton number of the final-state nucleus, $K_e$ is the electron's kinetic energy, $p_e$ is the electron's momentum, $E_0$ is the Q-value of the decay, and $U_{ei}$ are the elements of the PMNS matrix for neutrino mass states $m_i$, $i=1-3$.  The only dependence on the neutrino mass comes from the phase-space factor.  It is independent of all other properties of the neutrino, including whether neutrinos are Majorana or Dirac particles.

Two primary techniques have been considered for measuring the neutrino mass with beta decays.  Cryogenic bolometers are being employed by the MARE collaboration~\cite{ref:mare} to measure the heat deposited in the calorimeters by the beta decays of Re.  The first iteration of the experiment aims to confirm the measurements made by the Mainz and Troitsk experiments and place a limit at the eV level.  The follow-up experiment, MARE-II, will reach for the sub-eV level.

The second technique involves the use of a spectrometer to precisely select high-energy electrons from tritium decays.  The most recent experiments to use this technique are the Mainz and Troitsk experiments.  They placed similar limits on the neutrino mass: $m_{\beta\nu} < 2.3\ \mathrm{eV}$~\cite{ref:mainz,ref:troitsk}.  KATRIN, aims to lower that limit by an order of magnitude, to 200~meV (90\% CL)~\cite{ref:katrin,ref:katrin2}.  KATRIN is currently under construction in Karlsruhe, Germany.  In aiming to bring the mass limit down to 200~meV, KATRIN will have reached the practical limits of the spectrometer method.  The new technique for determining the electron energy, using cyclotron radiation, is described in the following section.

\section{A New Technique}

Tritium has several advantages with respect to its use as a source for measuring the neutrino mass.  It is a simple nucleus, so the matrix element of the decay is relatively easy to calculate.  The Q-value for the reaction is relatively low, at 18.575~eV, and the half-life is only 12.3~years.  Since the spectrometer technique is reaching its practical technological limit, it is worth investigating new techniques that can further take advantage of tritium's convenient properties.

The Project 8 collaboration proposes an alternate method of measuring the electron energies: measure the cyclotron radiation emitted by the electrons spiraling around magnetic field lines.  An enclosed volume of tritium is placed in a uniform magnetic field.  As the tritium nuclei decay, the electrons will spiral around the magnetic fields lines.  As they do so, they are being accelerated, and therefore emit cyclotron radiation.  The frequency of that radiation is proportional to the magnetic field strength, and inversely proportional to the electron's kinetic energy:

\begin{equation}
\omega = \frac{eB}{\gamma m_e} = \frac{\omega_c}{\gamma} = \frac{\omega_c}{1+K_e/(m_e c^2)}.
\label{eq:frequency}
\end{equation}

By measuring the frequency of the cyclotron radiation, one can measure the electron's kinetic energy without interfering with the electron itself.  Using a 1-T magnetic field, the endpoint of the tritium spectrum falls around 26~GHz.  This is a frequency band of significant commercial interest; we can therefore take advantage of commercially-available antennas and detectors.

The power emitted as cyclotron radiation depends both on the relativistic velocity electron, $\beta$, and the angle at which it is emitted relative to the direction of the magnetic field, $\theta$:

\begin{equation}
P(\beta,\theta) = \frac{1}{4\pi\epsilon_0}\frac{2 q^2 \omega_c^2}{3c}\frac{\beta_{\perp}^2}{1-\beta^2}, \qquad \beta_{\perp} \equiv \beta \sin{\theta}.
\label{eq:power}
\end{equation}

The electrons that radiate the most power will be the easiest to detect, because the signal-to-noise ratio will be higher.  Equation~\ref{eq:power} shows that the power will be greatest for electrons with $\theta \approx 90^{\circ}$.  Conveniently, these electrons also travel the slowest along the magnetic field lines, increasing the amount of time they can be observed.

\begin{figure}
\includegraphics[width=5in]{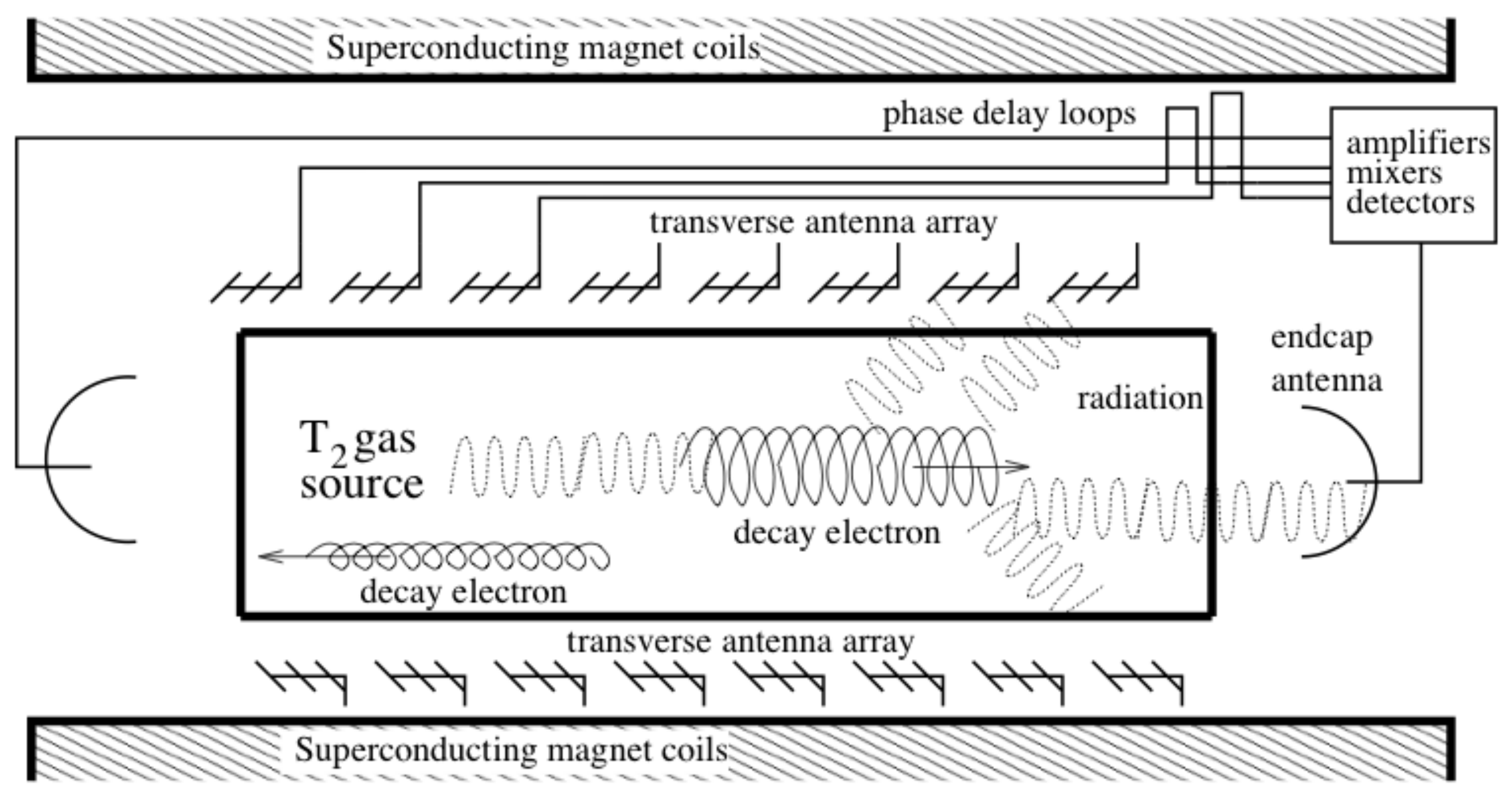}
\caption{\label{fig:simulationsetup}One possible setup for an experiment measuring the cyclotron radiation from tritium beta decay electrons in a magnetic field.  This design was used in an initial simulation to examine the potential for this type of experiment~\cite{ref:formaggio_monreal_2009}.  The tritium gas is enclosed in a chamber that is transparent to microwaves.  The chamber is then surrounded by an antenna array, all of which is placed inside the bore of a 1~T magnet.}
\end{figure}

Several configurations for the experiment are under consideration.  Figure~\ref{fig:simulationsetup} shows one possible setup.  The tritium is contained within a chamber inside of the bore of a 1-T magnet.  An antenna array is arranged around the outside of the source chamber.  As the electrons radiate, the microwave radiation passes through the chamber wall and is detected by the antenna array.  The signals from the individual antennas in the array can be summed coherently, since multiple antennas will be viewing each electron at the same time.

\begin{figure}
\includegraphics[width=5in]{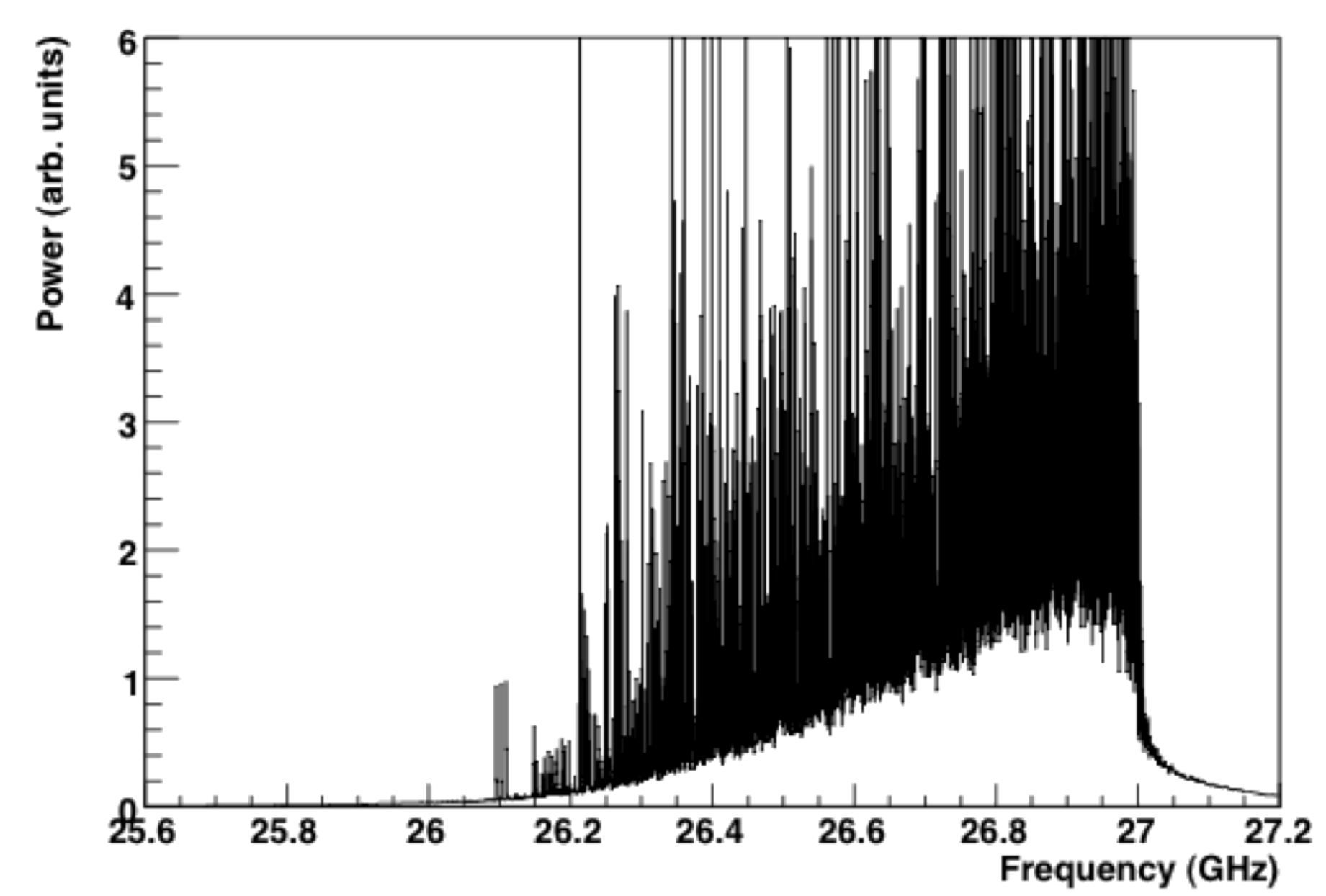}
\caption{\label{fig:powerspectrum}Simulated power spectrum from the hypothetical experiment shown in Fig.~\ref{fig:simulationsetup}.  $10^5$ beta decays were simulated over 30~$\mu$s in a 10-m-long uniform magnet ($B=1\ \mathrm{T}$).  The vertical arrow indicates the location of the tritium beta-decay endpoint in frequency space.}
\end{figure}

Figure~\ref{fig:powerspectrum} shows the power spectrum from a simulation of 100,000 tritium decays in 30~$\mu$s in the hypothetical experiment in Fig.~\ref{fig:simulationsetup}.  Since frequency is inversely proportional to electron energy, the rare high-energy electrons are at lower frequency, near 26~GHz.  Low-energy electrons, making up the vast majority of the spectrum, are piled up towards 27~GHz.  The pileup from low-energy electrons is a possible complication: unlike spectrometer-type experiments, we do not have an intrinsic method for rejecting low-energy electrons.  Instead, we can narrow the bandwidth such that the event rate is low enough that individual events can be identified.

The primary concern for making a precise electron energy measurement is the ability to measure frequency precisely.  The desired energy precision is therefore the place to start in considering the requirements for this type of experiment.  To achieve the necessary energy precision, $\Delta E$, we need to achieve a relative frequency precision of $\Delta f / f = \Delta E / m_e$.  KATRIN is designed to achieve $\Delta E \approx 1\ \mathrm{eV}$; for Project 8 to achieve a similar accuracy means that $\Delta f / f \approx 2 \times 10^{-6}$.  This accuracy is reasonable with current technologies.  With a 1-T magnetic field, $\Delta f \approx 52\ \mathrm{kHz}$ at 26~GHz.

The desired frequency accuracy determines for how long we must be able to observe single electrons.  To have a frequency resolution of $\Delta f$, we must measure each electron for $t_{\mathrm{min}} = 1 / \Delta f$.  With the design parameters discussed above, the electrons must be coherently measured for at least 20~$\mu$s.  The minimum measurement time places constraints on a number of physical parameters of the experiment.  The gas density must be low enough that, on average, 18.6~keV electrons can travel for $t_{\mathrm{min}}$ without scattering.  Furthermore, the experiment must be large enough so that the electron can be tracked continuously.

\begin{figure}
\includegraphics[width=5in]{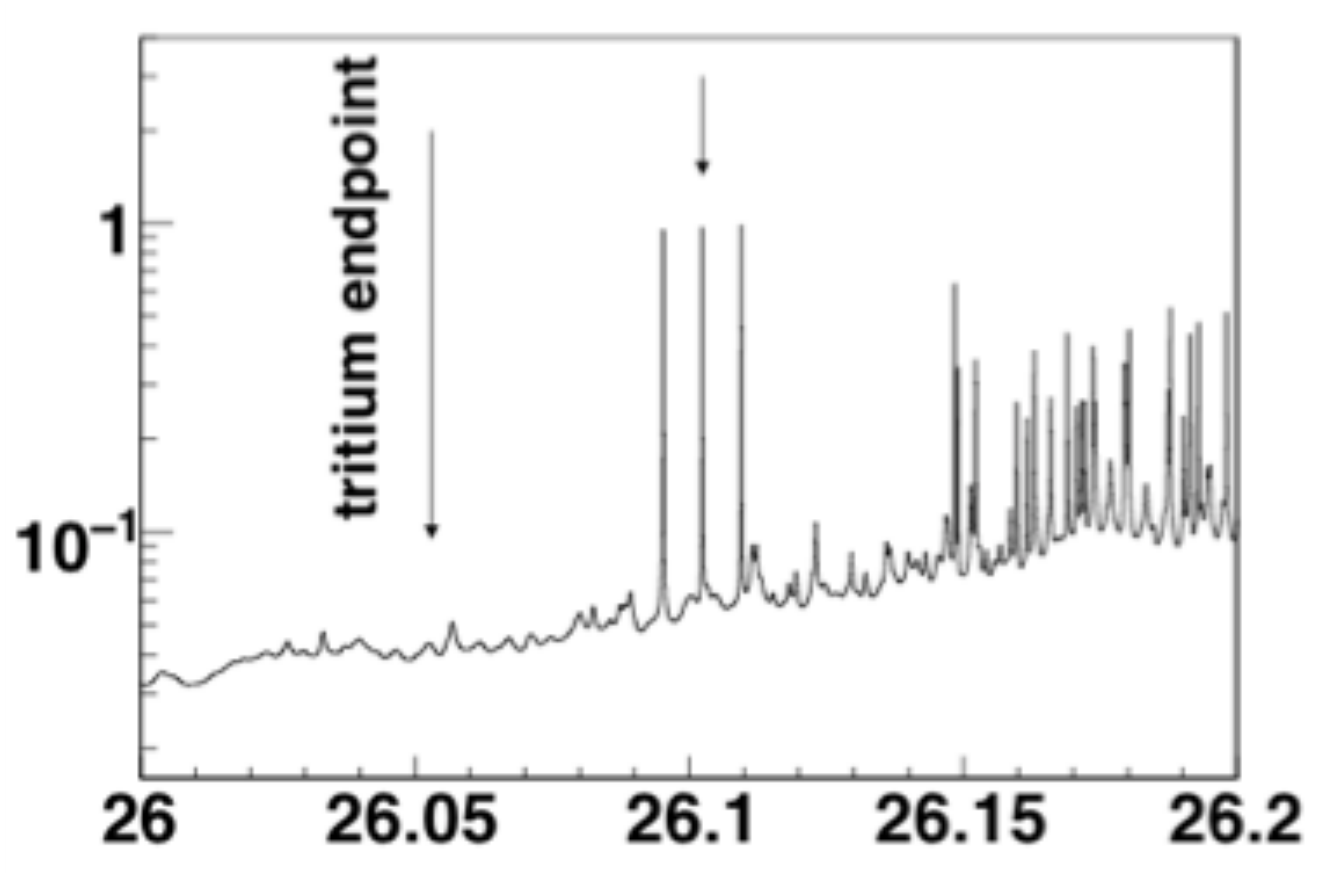}
\caption{\label{fig:zoomedpowerspectrum}Zooming in on the beta-decay endpoint of the power spectrum from Fig.~\ref{fig:powerspectrum}.  The triplet of peaks from a high-energy electron are easily distinguished from the background.}
\end{figure}

The signal detected for a single electron may be more complicated than the single frequency at which the cyclotron radiation is emitted.  In particular, the detected signal can include a Doppler shift due to the velocity of the electron parallel to the magnetic field, $\beta_{\parallel}$, a dependence on the electron-antenna distance, and effects from the angular dependence of the power distribution of the radiation.  The way these effects are represented in the data depends strongly on the antenna configuration.  For the hypothetical experiment shown in Fig.~\ref{fig:simulationsetup}, if the signals from the different antennas are summed coherently, there will be sideband peaks from the Doppler shift.  Fig.~\ref{fig:zoomedpowerspectrum} zooms in on the high-energy (low-frequency) region of the power spectrum shown previously.  In this simulation a triplet of peaks from a single high-energy electron are easily distinguishable.  Though it is an antenna-design-dependent effect, the triplet of peaks due to the Doppler shift could be a convenient tool for tagging electrons.

\subsection{Prototype Experiment}

The Project 8 Collaboration has put together a prototype experiment to explore the practical use of cyclotron radiation as a method for measuring the decay-electron energy.  The initial goal of the prototype is to verify that we can, in fact, detect the cyclotron radiation from a single electron.  We will use a $^{83m}$Kr radioactive source, which emits a monoenergetic electron.  This excited nucleus emits a 17.8~keV electron, and has a half-life of 1.83 hours.  The source is a good stand-in for tritium: it is gaseous, emitting the electrons isotropically, and the electron energy is close to the tritium-decay endpoint.

\begin{figure}
\includegraphics[width=3in]{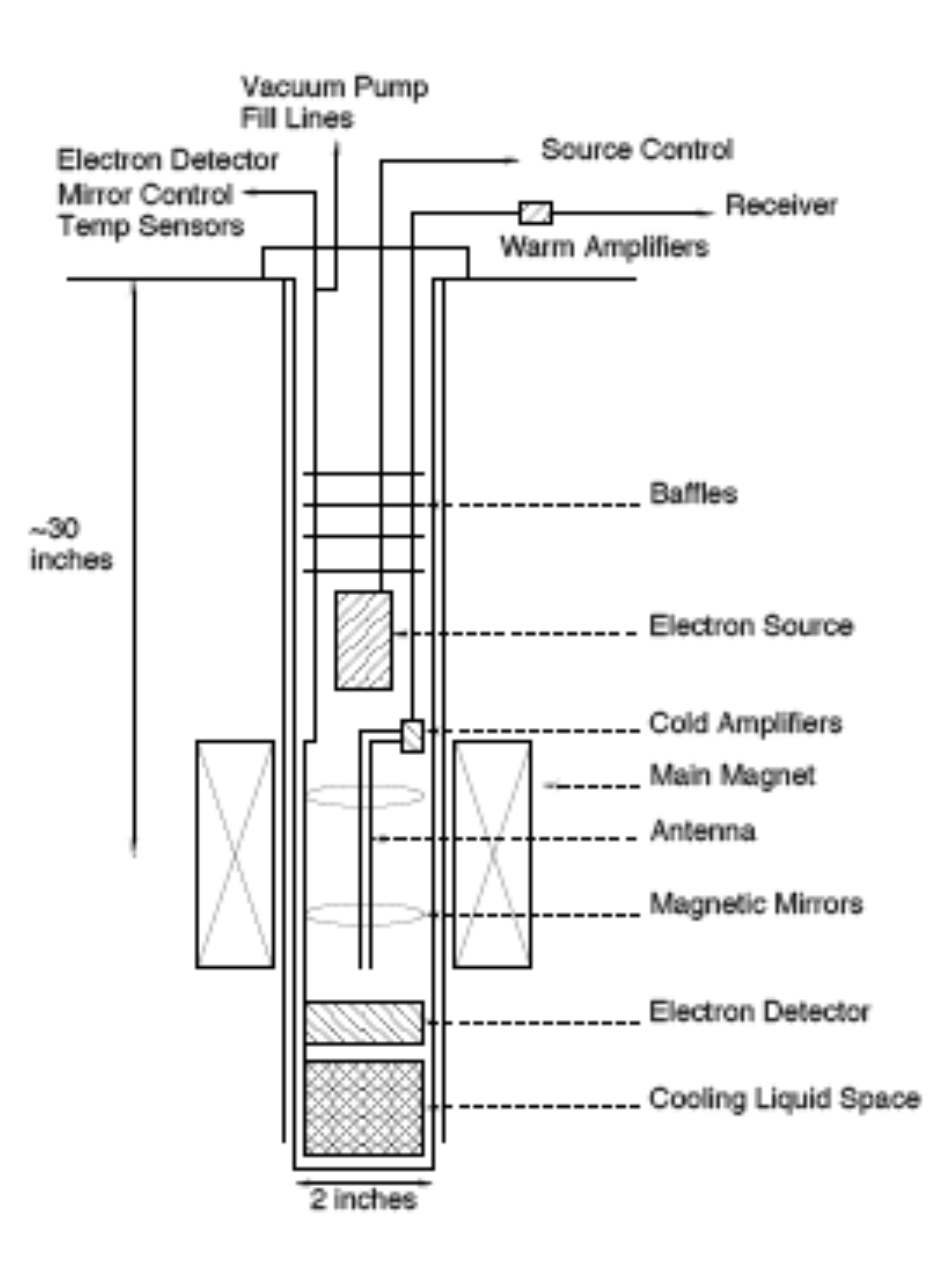}
\caption{\label{fig:prototype}Diagram of the prototype experiment located at the University of Washington, in Seattle, WA.  This particular design shows a two-wire antenna to detect the cyclotron radiation, though other designs are also being considered and tested.}
\end{figure}

Figure~\ref{fig:prototype} shows a diagram of the prototype, which is located at the University of Washington, in Seattle, WA.  A superconducting solenoid provides the 1-T magnetic field.  The electrons are trapped in a small ($\approx 1 \mathrm{mm}^3$) magnetic bottle in the bore of the magnet.  The magnetic field traps the electrons in the horizontal plane; a current loop within the bore of the magnet decreases the field slightly in a small volume, trapping the electrons vertically.  Whether or not electrons are trapped depends on the depth of the magnetic bottle potential, and the pitch angle of the electrons.  Electrons with large $\beta_{\perp}$ will be trapped.  Fortunately, these electrons also emit the most power as cyclotron radiation.  For the initial setup of the prototype, only electrons with large pitch angles ($\theta \ge 89^{\circ}$) are trapped.  Though this angle selection severely limits the number of electrons we will detect, it maximizes the signal-to-noise ratio.

The ability to open the magnetic bottle by turning off or reversing the current in the current loop will allow us to confirm that we are indeed trapping electrons, and accurately measure the noise levels. In addition to detecting the cyclotron radiation, we will employ more traditional means of detecting electrons to monitor the presence of $^{83m}$Kr in the trap, and verify that electrons are actually trapped.

Multiple antenna configurations will be investigated.  The diagram in Fig.~\ref{fig:prototype} shows a two-wire antenna.  The two wires act as a waveguide.  The cyclotron radiation couples to the transmission modes of the waveguide, and is propagated to the ends.  One end is attached to a low-noise cryogenic amplifier, after which the signal is transmitted out of the magnet bore, and mixed to a lower frequency.  A second antenna design being considered is a rectangular waveguide.  This design is particularly attractive because the gas can be confined within the waveguide horizontally, and with Kapton windows in the vertical direction.  The rectangular-waveguide antenna is capable of detecting signals emitted anywhere within the gas volume, whereas the two-wire antenna is most sensitive to electrons between the wires.

Once we have shown that we can detect single electrons using their cyclotron radiation, we will investigate the energy resolution achievable with this type of setup.  The $^{83m}$Kr source is particularly useful for this, since the electrons are monoenergetic.  We plan to use a tunable microwave source to perform power-vs.-frequency calibrations.

Finally, we want to demonstrate that the signal from cyclotron radiation can be used to identify electrons and determine their energy without additional detection methods.  Though for the initial stages of the prototype the data acquisition is untriggered, we will need to develop the ability to recognize electrons and trigger the recording of data.

At the time of this writing, the prototype has successfully collected an initial set of data using the rectangular waveguide antenna.  Analysis of the data is ongoing, and more data will be collected in the near future.  

\section{Summary}

Tritium beta decay experiments are currently the best method for directly probing the neutrino mass.  The KATRIN experiment has a sensitivity of 200~meV, but it has reached the practical technological limits for spectrometer-type experiments.  If we want to examine below the hundred-meV range, we need a new technique.  Project 8 aims to use radiofrequency techniques to measure the beta-decay electron energy.  A prototype experiment is currently in use to examine the feasibility of the technique, and to provide some initial insight into the possible sensitivity that might be achieved.


\bigskip 
\end{document}